\documentstyle[aps]{revtex}
\begin{document}
\baselineskip=25pt
\draft
\preprint
\title{ ELECTRODYNAMICS CLASSICAL INCONSISTENCIES}
\author{Manoelito M de Souza}
\address{Universidade Federal do Esp\'{\i}rito Santo\\Departamento de
F\'{\i}sica\\29065.900 -Vit\'oria-ES-Brasil}
\thanks{e-mail:manoelit@cce.ufes.br }
\date{\today}
\maketitle
\begin{abstract}
\noindent The problems of Classical Electrodynamics with the electron equation
of motion and with non-integrable singularity of its self-field  stress tensor
are well known. They are consequences, we show, of neglecting terms that are
null off the charge world line but that gives a non null contribution on its
world line. The self-field stress tensor of a point classical electron is
integrable, there is  no causality violation and no conflict with energy
conservation in its equation of motion, and there is no need of any kind of
renormalization nor of any change in the Maxwell's theory for this.\\ (This is
part of the paper hep-th/9510160, stripped , for simplicity, of its
non-Minkowskian geometrization of causality and of its discussion about the
physical meaning of the Maxwell-Faraday concept of field).

\end{abstract}

Three unsolved problems \cite{Rohrlich,Jackson} make of Classical
Electrodynamics of a point electron a non-consistent theory: 1) the field
singularity in the Lienard-Wiechert solution; 2) the non-integrable
singularities of its stress tensor; 3) the bizarre, causality violating
behaviour of solutions of the Lorentz-Dirac equation \cite{Rowe,Lozada}.\\
The Lorentz-Dirac equation,
\begin{equation}
\label{lordireq}
ma = e F_{ext} .V +\frac{2e^{2}}{3}(\stackrel{.}{a}-a^{2}V),
\end{equation}
is the greatest paradox of classical field theory as it cannot simultaneously
preserve both the causality and the energy conservation, although there is
nothing in the premisses for its derivation that justify such violations.
The presence of the Schott term, $\frac{2e^{2}}{3}\!\stackrel{.}{a}$, is  the
cause  of  all pathological features  of (\ref{lordireq}), like microscopic
non-causality,  runaway solutions, preacceleration, and other bizarre effects
\cite{Eliezer}. On the other hand the  presence of this term  is necessary for
the  maintenance of energy-momentum  conservation; without it it would be
required a null radiance for an accelerated charge, as
$\stackrel{.}{a}.V+a^{2}=0$. The argument, although correct, that such
causality violations are not observable because they are outside the scope of
classical physics \cite{Jackson} and are blurred \cite{Zuber}  by
quantum-mechanics effects is not enough compelling,  because these same
problems  remain in a quantum formalism, just disguised in other apparently
distinct problems \cite{hep-th/9505169}.\\
The presence of non-integrable singularities in the electron self-field stress
tensor is also a major problem. Previous attempts on taming these singularities
have relied on modifications of the Maxwell's
theory with ad hoc addition of extra- terms to the field stress tensor on the
electron world-line (see for example the reviews
\cite{Rowe,Lozada,Teitelboim}); it is particularly interesting that, as we will
show
here, instead of adding anything we should actually not drop out some null
terms. Is their
contribution (not null, in an appropriate limit) that avoid the infinities.
The same problem happens in the derivations of the electron equation of motion:
they are done with incomplete field expressions that do not contain these terms
that are null only off
the particle world-line. The Schott term in the Lorentz-Dirac equation is a
consequence of this; it does not
appear in the equation when the full field expression is correctly used.\\
The solution to the first problem, which is in essence the origin of the second
and third problems, involves a discussion of the meaning of the Maxwell-Faraday
concept of field and requires an entirely new formalism  based on an
anticipated recognition, yet in  a classical context, of the discrete and
localized character of the electromagnetic interaction, which is proper of a
quantum formalism. The solution to the second and to the third problems is
easier to describe and just requires the knowledge of the model of spacetime
geometry behind this new formalism. We will present here a simplified
Minkowskian version of it. The complete Riemannian version and its relevance to
Quantum Field Theory and to the question of the origin and meaning of mass, is
too long to be included in this letter. It is being discussed elsewhere
\cite{hep-th/9510160,hadrons}. The immediate difference in the outcomes of
these two versions appears in the equation of motion for a point electron,
\begin{equation}
\label{anticipation}
m\hbox{\Large a}^{\mu}-<\frac{1}{4}\;{\hbox{\Large a}}^{2}\hbox{\Large
a}^{\mu}>=F^{\mu}_{ext}-<\frac{2{\hbox{\Large a}}^{2}}{3}V^{\mu}>.
\end{equation}
The last term on the LHS corresponds to the energy associated to the Riemannian
curvature of the electron manifold and it is not present in its Minkowskian
counterpart (\ref{ffinally}). This is not the Lorentz-Dirac equation as it does
not contain the troublemaker third-order term. So, it does not violate
causality, and it is, nonetheless, compatible with 4-momentum conservation, as
we discuss below.\\
Equation (\ref{lordireq}) can be obtained from  energy-momentum conservation
in the Lienard - Wiechert solution \cite{Rohrlich,Rowe,Lozada,Teitelboim},
\begin{equation}
\label{1} A=\frac{V}{\rho}{\Bigg|}_{{\tau}_{ret}},\;\;\;for\;\;\;\rho>0,
\end{equation}
in the   limit of $\rho\rightarrow0$.  V is the tangent to the particle
world-line $z=z(\tau)$, parameterized  by its proper time $\tau$,
($V=dz/d\tau$, and $V^{2}=-1$). It is expressed in terms of retarded
coordinates \cite{Rohrlich,Teitelboim}, by which any spacetime point x is
constrained  with  a particle world-line point $z(\tau)$ by
$$R^{2}=0,\;\;\;\;\ R^{0}>0,$$
and
\begin{equation}
\label{dlcone}
d\tau+K.dx=0,
\end{equation}
where $R\equiv x-z(\tau), \;  \;\;\rho\equiv -V.\eta.R$, where $\eta$ is the
Minkowski metric tensor, (with signature +2).  $\rho$ is the spatial distance
between the point  x where the electromagnetic field is observed and the point
$z(\tau)$, position of the charge, in  the charge rest frame at its retarded
time. $R^{2}=0$ implies on $\rho=\Delta\tau$.  K, defined by
\begin{equation}
\label{K}
K^{\mu}:=\frac{\Delta x^\mu}{\rho}=\frac{\Delta x^\mu}{\Delta\tau},
\end{equation}
is a null 4-vector, $K^{2}=0$, and represents a light-cone generator, or the
electromagnetic wave-front 4-vector.\\
The retarded Maxwell field, ${F_{\mu\nu}}_{ret}:=\partial_{[\nu}A_{\mu]}$, is
given by
\begin{equation}
\label{Fl}
{F_{\mu\nu}}_{ret}=\frac{1}{\rho^{2}}[K_{\mu},V_{\nu}+\rho(\hbox{\Large
a}_{\nu}+\hbox{\Large a}_{K}V_{\nu})]=\frac{1}{\rho}[K_{\mu},\hbox{\Large
a}_{\nu}]+\frac{\hbox{\Large
a}_{K}}{\rho}[K_{\mu},V_{\nu}]+\frac{[K_{\mu},V_{\nu}]}{\rho^{2}},
\end{equation}
where, for notational simplicity, we  are using  $\;(A,B):=AB+BA,\;$
$\;[A,B]:=AB-BA\;,$  ${\hbox{\Large a}}_K:=\hbox{\Large a}.K,$ and we are
making the electron charge and the speed of light equal to 1: $e=c=1$.\\
\noindent Using (\ref{Fl}) in
$4\pi\Theta_{\mu\nu}=F_{\mu\beta}\eta^{\alpha\beta}F_{\alpha\nu}-\eta_{\mu\nu}\frac{F^{\alpha\beta}F_{\alpha\beta}}{4},$ for finding the electron self-field stress tensor,
\begin{equation}
\label{t}
4\pi\rho^{4}\Theta=[K,\rho\hbox{\Large a}+V(1+\rho{\hbox{\Large
a}}_K)].\eta.[K,\rho\hbox{\Large a}+V(1+\rho{\hbox{\Large
a}}_K)]-\frac{\eta}{4}[K,\rho\hbox{\Large a}+V(1+\rho{\hbox{\Large a}}_K)]^{2},
\end{equation}
or $\Theta=\Theta_{2}+\Theta_{3}+\Theta_{4}$, with
\begin{equation}
\label{t2}
4\pi\rho^{2}\Theta_{2}=[K,\hbox{\Large a}+V{\hbox{\Large
a}}_K].\eta.[K,\hbox{\Large a}+V{\hbox{\Large
a}}_K]-\frac{\eta}{4}[K,\hbox{\Large a}+V{\hbox{\Large a}}_{K}]^{2},
\end{equation}
\begin{equation}
\label{t3}
4\pi\rho^{3}\Theta_{3}=[K,V].\eta.[K,\hbox{\Large a}+V{\hbox{\Large
a}}_K]+[K,\hbox{\Large a}+V{\hbox{\Large
a}}_K].\eta.[f,V]-\frac{\eta}{2}Tr[K,V].\eta.[K,\hbox{\Large a}],
\end{equation}
\begin{equation}
\label{t4}
4\pi\rho^{4}\Theta_{4}=[K,V].\eta.[K,V]-\frac{\eta}{2}[K,V]^{2}.
\end{equation}
 It is worth to explicitly write (\ref{t}-\ref{t4}) for $\rho>0$ (then
$K^{2}=0$) and make some comments.
\begin{equation}
\label{t2e}
4\pi\rho^{2}\Theta_{2\;\mu\nu}=-K_{\mu}K_{\nu}{\big\lbrace}{\hbox{\Large
a}}^{2}-{{\hbox{\Large a}}_{K}}^{2}{\big\rbrace},
\end{equation}
\begin{equation}
\label{t3e}
4\pi\rho^{3}\Theta_{3\;\mu\nu}=2K_{\mu}K_{\nu}{\hbox{\Large a}}_K -{\biggl (}
K_{\mu},({\hbox{\Large a}}+V{\hbox{\Large a}}_K)_{\nu}{\biggr )},
\end{equation}
\begin{equation}
\label{t4e}
4\pi\rho^{4}\Theta_{4\;\mu\nu}=K_{\mu}K_{\nu} -(K_{\mu},V_{\nu})
-\frac{\eta_{\mu\nu}}{2},
\end{equation}
$\Theta_{2}$, although singular at $\rho=0$, is nonetheless integrable. By that
it is meant that $\int d^{4}x\Theta_{2}$ exists \cite{Rowe}, while $\Theta_{3}$
and $\Theta_{4}$ are not integrable; they generate, respectively, the
problematic Schott term in the LDE and a divergent expression, the electron
bound 4-momentum \cite{Teitelboim}. The most update prescription
\cite{Rowe,Lozada} is to redefine $\Theta_{3}$ and $\Theta_{4}$ at the electron
world-line in order to make them integrable, but without changing them at
$\rho>0$, so to preserve the standard results of Classical Electrodynamics.
This is possible with the use of distribution theory, but it is always an
introduction of something strange to the system and in an {\it ad hoc} way. The
most unsatisfactory aspect of this procedure is that it regularizes the above
integral but leaves an unexplained and unphysical discontinuity in the flux of
4-momentum from the charge world-line: $\Theta(\rho=0)\neq\Theta(\rho\sim0).$
\\
We observe that
\begin{equation}
\label{C}
K_{\mu}\Theta_{2}^{\mu\nu}{\bigg|}_{\rho>0}=0,
\end{equation}
which is important in the identification of $\Theta_{2}$ with the radiated part
of $\Theta$, and that
\begin{equation}
\label{C1} K_{\mu}\Theta_{3}^{\mu\nu}{\bigg|}_{\rho>0}=0
\end{equation}
The important difference among the sets of equations (\ref{t2}-\ref{t4}) and
(\ref{t2e}-\ref{t4e}) is that while the equations in the first one are complete
, in the sense that they keep the  terms proportional to $K^{2}$, which are
null (as $K^{2}=0$), in the last set of equations they have been dropped off.
But these $K^{2}$-terms,  even with $K^{2}=0,$ should not be dropped from the
above equations, since they are necessary for producing the correct limits when
 $\rho\Longrightarrow0,$ or $x\Longrightarrow z.$ As
$K^{\mu}:=\frac{R^{\mu}}{\rho}$, in this limit  we have a $0/0$-type of
indeterminacy, which can be raised with the L'Hospital rule and
$\frac{\partial}{\partial\tau}.$  This results in
$$\lim_{R\to0}K=V,$$
and$$\lim_{R\to0}K_{\mu}K^{\mu}=\lim_{R\to0}\frac{R.\eta.R}{\rho^{2}}=-1.$$
A Feynmann diagram, see the figure 1, helps in the understanding of these two
results. In the limit of $\rho\rightarrow0,$ or at $\tau=\tau_{ret}$ there are
3 distinct velocities: K, the photon 4-velocity, and $V_{1}$ and $V_{2}$, the
electron initial and final 4-velocities. This is the reason for this
indeterminacy at $\tau=\tau_{ret}$. At $\tau=\tau_{ret}+d\tau$ there is only
$V_{2}$, and only $V_{1}$ at $\tau=\tau_{ret}-d\tau,$ or, back to the usual
picture, $V({\tau})$, in general. We must observe that the classical picture of
a continuous interaction cannot resolve this indeterminacy. The lesson one
should learn from this is that even in a classical context, is necessary to
take into account the discrete and localized  (quantum) character of the
fundamental electromagnetic interaction in order to have a clear and consistent
physical picture. This is the viewpoint adopted in
\cite{hep-th/9505169,hep-th/9510160}, where the concept of a classical photon
is introduced. It fulfils a requirement of an equal-foot treatment to the
electron and to its self electromagnetic field, which is more in accordance
with the experimental data that show both as equally fundamental physical
objects of Nature.\\
To find the limit of something when $\rho\rightarrow0$ will be done so many
times in this letter that it is better to do it in a more systematic way.   We
want to find
\begin{equation}
\label{LR}
\lim_{R\to0}\frac{N(R)}{\rho^{n}},
\end{equation}
where $N(R)$ is a homogeneous function of R,  $N(R){\Big|}_{R=0}=0$. Then, we
have to apply the L'Hospital rule consecutively until the indeterminacy is
resolved.  As $\frac{\partial\rho}{\partial\tau}=-(1+\hbox{\Large a}.R)$, the
denominator of (\ref{LR}) at $R=0$ will be different of zero only after the
$n^{th}$-application of the L'Hospital rule, and then, its value will be
$(-1)^{n}n!$\\
If p is the smallest integer such that $N(R)_{p}{\Big|}_{R=0}\ne0,$ where
$N(R)_{p}:=\frac{d^{p}}{d{\tau}^{p}}N(R)$, then
\begin{equation}
\label{NR}
\lim_{R\to0}\frac{N(R)}{\rho^{n}}=\cases{\infty,& if $p<n$\cr
              (-1)^{n}{\frac{N(0)_{p}}{n!}},& if $p=n$\cr
0,& if $p>n$\cr}
\end{equation}
\begin{itemize}
\item Example 1: $\cases{K=\frac{R}{\rho}.& $n=p=1 \Longrightarrow
\lim_{R\to0}K=V$\cr K^{2}=\frac{R.\eta.R}{\rho^{2}}.&$ n=p=2 \Longrightarrow
\lim_{R\to0}K^{2}=-1.$\cr}$
\item Example 2: $\frac{[K_{\mu},\hbox{\Large
a}_{\nu}]}{\rho}=\frac{[R_{\mu},\hbox{\Large a}_{\nu}]}{\rho^{2}}\;\;$
$\;\Longrightarrow \;$ $\;p=1<n=2\Longrightarrow
\lim_{R\to0}\frac{[K_{\mu},\hbox{\Large a}_{\nu}]}{\rho}$ diverge
\item Example 3: $\frac{{\hbox{\Large
a}}_K}{\rho}K^{[\mu}V^{\nu]}=-\frac{\hbox{\Large
a}.R}{\rho^{3}}R^{[\mu}V^{\nu]}\Longrightarrow p=4>n=3$ \qquad
$\lim_{\rho\to0}\frac{{\hbox{\Large a}}_K}{\rho}K^{[\mu}V^{\nu]}=0$
\item Example 4
$\frac{[K_{\mu},V_{\nu}]}{\rho^{2}}=\frac{[R_{\mu},V_{\nu}]}{\rho^{2}}\;\;
\;\Longrightarrow
\;\;p=2<n=3\Longrightarrow\lim_{R\to0}\frac{[K_{\mu},V_{\nu}]}{\rho^{2}}$
diverge
\end{itemize}
The second term of the RHS of (\ref{Fl}) does not contribute to the electron
self-field at $\rho=0,$  but the first and the third terms  diverge, as
expected, although they produce integrable contributions to $\Theta.$ Let us
find the integral of the electron self-field stress tensor at the electron
causality-line: $\lim_{\rho\to0}\int dx^{4}\Theta,$  or $\lim_{\rho\to0}\int
d\tau\rho^{2}d\rho d^{2}\Omega\Theta,$ in terms of retarded coordinates
\cite{Rowe,Synge,Tabensky} $x^{\mu}=z^{\mu}+\rho K^{\mu},$ where $d^{2}\Omega$
is the element of solid angle in the charge rest-frame.\\ But, we will first
prove a useful result to be used with (\ref{NR}), when the numerator has the
form $N_{0}=A_{0}.\eta.B_{0},$ where A and B represent two possibly distinct
homogeneous functions of R, and the subindices indicate the order of
$\frac{d}{d\tau}$. Then
\begin{equation}
\label{NP}
N_{p}=\sum_{a=0}^{p}\pmatrix{p\cr a\cr}A_{p-a}.\eta.B_{a}
\end{equation}
So, for using (\ref{NR}) with (\ref{NP}), we just have to find the
$\tau$-derivatives  of A and B that produce the first non- null term at the
limit of $R\rightarrow0$.\\
Applying (\ref{NR}) and (\ref{NP}) for finding $\lim_{\rho\to0}\int
dx^{4}\Theta$ we just have to consider the first term of the RHS of (\ref{t});
as the second one is the trace of the first, its behaviour under this limit can
be inferred.
\begin{itemize}
\item Example 5
$$\lim_{\rho\to0}\frac{\rho^{2}[K,\rho\hbox{\Large a}+V(1+\rho{\hbox{\Large
a}}_K)].\eta.[K,\rho\hbox{\Large a}+V(1+\rho\;{\hbox{\Large
a}}_K)]}{\rho^{4}}=$$
$$=\lim_{\rho\to0}\frac{[R,\rho\hbox{\Large a}+V(1+\hbox{\Large
a}.R)].\eta.[R,\rho\hbox{\Large a}+V(1+\hbox{\Large a}.R)]}{\rho^{4}}$$

So, $A_{0}=B_{0}=[R,\rho\;\hbox{\Large a}+V(1+\rho\hbox{\Large
a}.R)]\Longrightarrow A_{2}=B_{2}=[\hbox{\Large a},V]+{\cal O}(R).\;\;$
Therefore, according to (\ref{NP}), for producing a non null $N_{p}$, a and p
must be given by
$$p-a= a=2\Longrightarrow p=4=n\Longrightarrow N_{4}=6[\hbox{\Large
a},V].\eta.[\hbox{\Large a},V]+{\cal O}(R).$$
Then, we conclude from (\ref{NR}), that

$$\lim_{\rho\to0}\frac{\rho^{2}[K,\rho\hbox{\Large a}+V(1+\rho{\hbox{\Large
a}}_K)].\eta.[K,\rho\hbox{\Large a}+V(1+\rho{\hbox{\Large
a}}_K)]}{\rho^{4}}=\frac{1}{4}[\hbox{\Large a},V].\eta.[\hbox{\Large a},V]$$
We have, therefore, from (\ref{t}), that
\begin{equation}
\label{te}
\lim_{\rho\to0}\int dx^{4}\;\Theta=[\hbox{\Large a},V].\eta.[\hbox{\Large
a},V]-\frac{\eta}{4}[\hbox{\Large a},V]^{2}
\end{equation}
The flux of 4-momentum irradiated from the electron, which is the meaning of
(\ref{te}), is finite and depends only on its instantaneous velocity and
acceleration.
It is interesting that (\ref{te}) comes entirely from the velocity term,
$\frac{[K,V]}{\rho^{2}}$, as we can see from the following example.
\item Example 6
$$\lim_{\rho\to0}\frac{\rho^{2}[K,V].\eta.[K,V]}{\rho^{4}}=\lim_{\rho\to0}\frac{[R,V].\eta.[R,V]}{\rho^{4}}=\frac{1}{4}[\hbox{\Large a},V].\eta.[\hbox{\Large a},V]$$
as $A_{2}=B_{2}=[\hbox{\Large a},V]+[R,\dot {\hbox{\Large a}}]\Longrightarrow
p-a=a=2\Longrightarrow N_{4}=6[\hbox{\Large a},V]+{\cal O}(R)\;\;$ and $
p=n=4.$ So,
$$\lim_{\rho\to0}\int dx^{4}\Theta=\lim_{\rho\to0}\int dx^{4}\Theta_{4}$$
The contribution from the other two terms just cancel to zero,
$$\lim_{\rho\to0}\int dx^{4}\Theta_{2}=-\lim_{\rho\to0}\int dx^{4}\Theta_{3},$$
as can be easily verified. We must realize that the discontinuity,
$$\Theta(\rho=0)\neq\Theta(\rho\sim0),$$ still remains. It is a consequence of
the problem 1; its solution requires an understanding of the physical meaning
of the Maxwell-Faraday concept of field in this context. It is being discussed
elsewhere \cite{hep-th/9510160}.
\end{itemize}
\begin{center}
The electron equation of motion
\end{center}
Let us now discuss the third problem. The electron equation of motion,  can be
obtained from
$$\lim_{\varepsilon\to0}\int
dx^{4}\partial_{\nu}T^{\mu\nu}\theta(\rho-\varepsilon)=0,$$
where $T^{\mu\nu}$ is the total electron energy-momentum tensor, which includes
the contribution from the electron kinetic energy, from its interaction with
external fields and from its self field. Let us move directly to the part that
will produce novel results:
\begin{equation}
\label{eqm}
m\int a^{\mu}d\tau=\int F_{ext}^{\mu}d\tau-\lim_{\varepsilon\to0}\int
dx^{4}\partial_{\nu}\Theta^{\mu\nu}\theta(\rho-\varepsilon),
\end{equation}
\noindent where $F_{ext}^{\mu}$ is the external forces acting on the electron,
and the last term represents the impulse carried out by the emitted
electromagnetic field in the Bhabha tube surrounding the electron world-line,
defined by the Heaviside function, $\theta(\rho-\varepsilon)$.  Using  the
divergence theorem, we have that the last term of the RHS of (\ref{eqm}) is
transformed into
\begin{equation}
\label{gauss}
\lim_{\varepsilon\to0}\int
dx^{4}\Theta^{\mu\nu}\partial_{\nu}\rho\;\delta(\rho-\varepsilon).
\end{equation}
\noindent This term represents  the flux of 4-momentum through the cylindrical
hypersurface $\rho=\varepsilon.$ Let us denote it by $P^{\mu}$:
\begin{equation}
\label{P}
P^{\mu}=\lim_{\varepsilon\to0}\int
dx^{4}\Theta^{\mu\nu}\partial_{\nu}\rho\;\delta(\rho-\varepsilon),
\end{equation}
As $\partial_{\nu}\rho= \rho {\hbox{\Large a}}_KK_{\nu}+K_{\nu}-V_{\nu},$ and
$\Theta=\Theta_{2}+\Theta_{3}+\Theta_{4},$ we can write
$P^{\mu}:=P^{\mu}_{0}+P^{\mu}_{1}+P^{\mu}_{2},$ with
\begin{equation}
\label{P2}
P^{\mu}_{2}=\lim_{\varepsilon\to0}\int
dx^{4}\Theta^{\mu\nu}_{4}(K-V)_{\nu}\;\delta(\rho-\varepsilon),
\end{equation}
\begin{equation}
\label{P1}
P^{\mu}_{1}=\lim_{\varepsilon\to0}\int dx^{4}\lbrace
\Theta^{\mu\nu}_{4}K_{\nu}\;\rho\;{\hbox{\Large
a}}_K+\Theta^{\mu\nu}_{3}(K-V)_{\nu}\rbrace\;\delta(\varepsilon-\rho),
\end{equation}
\begin{equation}
\label{P0}
P^{\mu}_{0}=\lim_{\varepsilon\to0}\int dx^{4}\lbrace
\Theta^{\mu\nu}_{3}K_{\nu}\rho\hbox{\Large
a}_{K}+\Theta^{\mu\nu}_{2}(K-V)_{\nu}\rbrace\;\delta(\rho-\varepsilon),
\end{equation}
\noindent $P^{\mu}_{1}$ and $P^{\mu}_{2}$ are both null. In order to show this
we need to apply (\ref{NR}) for a N(R) with a generic form $N=A.\eta.B.C$,
where A,B, and C are  functions of R, such that $N(R=0)=0$. Then it is easy to
show, from (\ref{NP}), that
\begin{equation}
\label{AgBC}
N_{p}=\sum_{a=0}^{p}\sum_{c=0}^{a}\pmatrix{p\cr a\cr}\pmatrix{a\cr
c\cr}A_{p-a}.\;\eta\;.B_{a-c}.C_{c}
\end{equation}
\begin{itemize}
\item From (\ref{t4}), the integrand of (\ref{P2}), produces (again, we do not
need to consider the trace term)
$$\lim_{\rho\to0}\frac{\rho^{2}[K,V].\eta.[K,V].(K-V)}{\rho^{4}}=\lim_{\rho\to0}\frac{[R,V].\eta.[R,V].(R-\rho V)}{\rho^{5}},$$
or, schematically
$$\lim_{\rho\to0}\frac{A.\eta.A.C}{\rho^{5}}$$
with $A_{0}=B_{0}=[R,V],$ and $C_{0}=(R-V\rho).$ Then, $A_{2}=[a,V]+{\cal
O(R)},\;\;$   $\;C_{2}=a+{\cal O(R)},$ and we have, from (\ref{AgBC}),  the
following restrictions on a and c for producing a $N(R=0)_{p}\ne0:$

$c=2;\quad  a-c=2;$ and $p-a=2$ or $p=6>n=5. \quad$
Therefore, according to (\ref{NR})
\begin{equation}
\label{P20}
P^{\mu}_{2}=0
\end{equation}
\item From the second term of the integrand of (\ref{P1}) and from (\ref{t3})
we have
$$\lim_{\rho\to0}\frac{\rho^{2}[K,V].\eta.[K,\hbox{\Large a}+V{\hbox{\Large
a}}_K].(K-V)}{\rho^{3}}=\lim_{\rho\to0}\frac{[R,V].\eta.[R,\rho\;\hbox{\Large
a}+V\hbox{\Large a}.R].(R-\rho\;V)}{\rho^{5}},$$
and then, from (\ref{AgBC}) with
 $$C_{0}=R-V\rho\;\;\Longrightarrow C_{2}={\hbox{\Large a}}+{\cal
O(R)}\Longrightarrow c=2.$$
$$B_{0}=[R,V]\;\;\Longrightarrow B_{2}=[\hbox{\Large a},V]+{\cal
O(R)}\Longrightarrow a=4.$$
$$A_{0}=[R,{\hbox{\Large a}}]\;\;\Longrightarrow A_{1}=[{\hbox{\Large
a}},V]+{\cal O(R)}\Longrightarrow p=5=n.$$
This produces $[\hbox{\Large a},V].\eta.[\hbox{\Large a},V].{\hbox{\Large
a}}={\hbox{\Large a}}^{2}{\hbox{\Large a}},$ which is cancelled by the equal
contribution from the trace term of (\ref{t3}).
Therefore, $$\lim_{\rho\to0}\rho^{2}\Theta^{\mu\nu}_{3}(K-V)_{\nu}=0$$
\item From the first term of the integrand of (\ref{P1}) and from (\ref{t4}) we
have
$$\lim_{\rho\to0}\frac{\rho^{2}[K,V].\eta.[K,V]K\rho{\hbox{\Large a}}.K
}{\rho^{4}}=\lim_{\rho\to0}\frac{[R,V].\eta.[R,V]R{\hbox{\Large a}}.R
}{\rho^{5}}.$$
Then, from (\ref{AgBC}), with $C=C_{0}=R\;\hbox{\Large a}.R\;\;\Longrightarrow
C_{3}=-3V{\hbox{\Large a}}^{2}+{{\cal O(R)}}\Longrightarrow c=3,$
and
$$A=B=B_{0}=[R,V]\;\;\Longrightarrow A_{2}= B_{2}=[\hbox{\Large a},V]+{{\cal
O(R)}}\Longrightarrow a=5 \hbox{ and }\;p=7>n=5.$$
Therefore, from (\ref{NR})
$$\lim_{\rho\to0}\rho^{2}\Theta^{\mu\nu}_{4}K_{\nu}\rho\;{\hbox{\Large
a}}_K=0$$
\end{itemize}
Consequently
 \begin{equation}
\label{P20}
P^{\mu}_{1}=0
\end{equation}
and
\begin{equation}
\label{P0F}
P^{\mu}=P^{\mu}_{0}
\end{equation}
\noindent $P^{\mu}_{0}$ is distinguished from  $P^{\mu}_{1}$ and $P^{\mu}_{2}$
for not being $\rho$-dependent. Therefore, it is not necessary to use the
L'Hospital rule on its determination, which, by the way,
 is not affected by the limit of $\varepsilon\rightarrow0$.
The physical meaning of this is that the flux of 4-momentum through the
cylindrical surface $\rho=\varepsilon$ comes entirely from the photon field.\\
 It is convenient now to introduce the spacelike 4-vector N, defined by
$N^{\mu}:=(K-V)^{\mu}$ and such that $N.\eta.V==0$ and $N.\eta.N=1$. N
satisfies
\begin{equation}
\label{n1}
\frac{1}{4\pi}\int d\Omega \overbrace{N\cdots N}^{odd\;number}=0
\end{equation}
\begin{equation}
\label{n2}
\frac{1}{4\pi}\int d\Omega NN=\frac{\eta+VV}{3};
\end{equation}
\noindent as can be found, for example, in references [4,13] or in the
appendix B of of reference [10]. From (\ref{t3}) and (\ref{C1}), or from
(\ref{n1}), we have
\begin{equation}
\label{t3f}
\int  d^{4}x\Theta^{\mu\nu}_{3}K_{\nu}\rho{\hbox{\Large a}}_{K}=0,
\end{equation}
{}From (\ref{t2e}), we have
\begin{equation}
4\pi\rho^{2}\Theta^{\mu\nu}_{2}N_{\nu}={\biggl (}({\hbox{\Large
a}.N})^{2}-{\hbox{\Large a}}^{2}{\biggr )}{\biggl (}V^{\mu}+N^{\mu}{\biggr )},
\end{equation}
which, with (\ref{n1}) and (\ref{n2}), gives
\begin{equation}
\label{tN}
\lim_{\rho\to0}\int  d^{4}x\Theta_{2}^{\mu\nu}N_{\nu}=-\int
d\tau\frac{2}{3}{\hbox{\Large a}}^{2}V^{\mu}.
\end{equation}
Then, from (\ref{tN}), (\ref{t3f}) and (\ref{P0F}), we have
\begin{equation}
\label{Larmor}
P^{\mu}=-\int d\tau\frac{2}{3}{\hbox{\Large a}}^{2}V^{\mu},
\end{equation}
the Larmor term.\\
Finally, from (\ref{eqm}), (\ref{gauss}), (\ref{P}) and (\ref{Larmor}),  we can
write the electron equation of motion, obtained from the Lienard-Wiechert
solution, as
\begin{equation}
\label{finally}
m\hbox{\Large a}^{\mu}=F^{\mu}_{ext}-\frac{2{\hbox{\Large a}}^{2}}{3}V^{\mu},
\end{equation}
 The external force provides the work for changing the charge velocity and for
the energy dissipated by the radiation.
It is non-linear, like the Lorentz-Dirac equation, but it does not contain the
Schott term, the responsible for its spuriously behaving solutions. This is
good since it signals that there will be no problem with causality violation.
But the Schott term in the Lorentz-Dirac equation has also the role of giving
the guaranty of energy conservation, which is obviously missing in
(\ref{finally}).  Assuming that the external force is of electromagnetic
origin, ($F_{ext}^{\mu}=F_{ext}^{\mu\nu}V_{\nu}),$ the contraction of V with
eq. (\ref{finally}) would require a contradictory $a^{2}\equiv0$. But this is
just an evidence that (\ref{finally}) cannot be regarded as a fundamental
equation. It would be better represented as
\begin{equation}
\label{ffinally}
m\hbox{\Large a}^{\mu}=F^{\mu}_{ext}-<\frac{2}{3}{\hbox{\Large a}}^{2}V^{\mu}>,
\end{equation}
with $$<\frac{2}{3}{\hbox{\Large
a}}^{2}V^{\mu}>=P^{\mu}=\lim_{\varepsilon\to0}\int
dx^{4}\Theta^{\mu\nu}\partial_{\nu}\rho\;\delta(\rho-\varepsilon),$$
It is just an effective or average result, in the sense that the contributions
from the electron self field must be calculated, as in (\ref{eqm}), by the
electromagnetic energy-momentum content of a spacetime volume containing the
charge world-line, in the limit of $\rho\rightarrow0$:
\begin{equation}
\label{ec}
m\int {\hbox{\Large a}}.Vd\tau=\int F_{ext}.Vd\tau-\lim_{\varepsilon\to0}\int
dx^{4}K_{\mu}\partial_{\nu}\Theta^{\mu\nu}\theta(\rho-\varepsilon).
\end{equation}
Observe that in the last term,V, the speed of the electron, is replaced by K
the speed of the electromagnetic interaction; only in the limit of
$R\rightarrow0$ is that $K\rightarrow V.$
We have to repeat the same steps from (\ref{eqm}) to (\ref{finally}) in order
to calculate this last term and to prove (done in the appendix) that it is
null:
 \begin{equation}
\label{fec}
\lim_{\varepsilon\to0}\int
dx^{4}K_{\mu}\partial_{\nu}\Theta^{\mu\nu}\theta(\rho-\varepsilon)=0,
\end{equation}
So, there is no contradiction anymore. Besides, it throws some light on the
physical meaning of $A_{\mu}$, which is discussed in \cite{hep-th/9510160}.

\section{\bf Appendix}

Let us prove (\ref{fec}). Its LHS implies on
\begin{equation}
\label{bfec}
-\lim_{\varepsilon\to0}\int
dx^{4}\Biggl\{(\partial_{\nu}K_{\alpha})\Theta^{\alpha\nu}\theta(\rho-\varepsilon)+K_{\mu}\Theta^{\mu\nu}\partial_{\nu}\rho\;\delta(\rho-\varepsilon)\Biggr\}.
\end{equation}

To find how the first term of the integrand of (\ref{bfec}) behaves in the
limit of R$\rightarrow0$, we need to find $\partial_{\nu}K_{\mu}$ from
$K^{\mu}=\frac{R^{\mu}}{\Delta\tau}$ and from  $\rho=\Delta\tau.$ The
difference between the derivatives of $\rho$ and of $\Delta\tau$ tends to zero
in the limit of $\rho\rightarrow0.$ So, it is irrelevant if we use one or the
other in the definition of K; we use the simplest one, $\Delta\tau,$ and
$\frac{\partial\Delta\tau}{\partial\tau}= -1$ to find:
$$\partial_{\nu}K_{\mu}=\frac{1}{\Delta\tau}{\Big\lbrace}\eta_{\mu\nu}+K_{\nu}(V-K)_{\mu}){\Big\rbrace}=\frac{1}{(\Delta\tau)^{3}}{\Big\lbrace}(\Delta\tau)^{2}\eta_{\mu\nu}+\Delta\tau V_{\mu}R_{\nu}-R_{\mu}R_{\nu}{\Big\rbrace}$$
Then, the first term in the integrand of (\ref{bfec}) contains
\begin{equation}
\label{cfec}
\lim_{R\to0}\frac{[R,\rho\;{\hbox{\Large a}}+V(1+{\hbox{\Large
a}}.R)].\eta.[R,\rho{\hbox{\Large a}}+V(1+{\hbox{\Large
a}}.R)]{\Big\lbrace}(\Delta\tau)^{2}\eta_{\mu\nu}+\Delta\tau
V_{\mu}R_{\nu}-R_{\mu}R_{\nu}{\Big\rbrace}}{\rho^{7}},
\end{equation}
and, using the notation of (\ref{AgBC}),
$$A_{0}=B_{0}=[R,\rho\;\hbox{\Large a}+V(1+\rho\hbox{\Large
a}.R)]\Longrightarrow A_{2}=B_{2}=[\hbox{\Large a},V]+{\cal
O}(R)\Longrightarrow a=4$$
$$C_{0}=(\Delta\tau)^{2}\eta_{\mu\nu}+\Delta\tau
V_{\mu}R_{\nu}-R_{\mu}R_{\nu}\Longrightarrow
C_{2}=2\eta+{\cal O}(R)\Longrightarrow c=2\Longrightarrow p=6<n=7$$
According to (\ref{NR}), this would produce a divergent result if $N_{6}\neq0,$
but $\Theta$ and its limit are traceless ($\Theta^{\mu\nu}\eta_{\mu\nu}=0)$,
and so, $N_{6}=0.$ Therefore, the indeterminacy $0/0$ remains and a new
application of the L'Hospital rule is demanded. Then, from (\ref{NP}), for
$p=6,\;a=4,\;c=2$, we have
$$N_{6}=\pmatrix{6\cr 2\cr}\pmatrix{2\cr 2\cr}A_{2}.\;\eta\;.B_{2}.C_{2},$$
and then
$$N_{7}=\dot{N}_{6}=\pmatrix{6\cr 2\cr}\pmatrix{2\cr
2\cr}{\Big\lbrace}A_{3}.\;\eta\;.A_{2}C_{2}+A_{2}.\;\eta\;.A_{2}C_{2}+A_{2}.\;\eta\;.A_{3}C_{2}+A_{2}.\;\eta\;.A_{2}C_{3}{\Big\rbrace}.$$
The terms containing $C_{2}$ will still give a null contribution ($\Theta$ is
traceless).
With $C_{3}=\frac{3}{2}(V,\hbox{\Large a})+{\cal O}(R),$
\begin{equation}
\label{dfT3}
{\Big\lbrace}A_{2}.\;\eta\;.A_{2}.C_{3}{\Big\rbrace}={\Big\lbrace}[V,\hbox{\Large a}].\eta.[V,\hbox{\Large a}].(9V\hbox{\Large a}+6\hbox{\Large a}V){\Big\rbrace}\equiv0,
\end{equation}
and then,
\begin{equation}
\label{dfT}
\lim_{\varepsilon\to0}\int
dx^{4}\partial_{\nu}K_{\alpha}\Theta^{\alpha\nu}\theta(\rho-\varepsilon)=0.
\end{equation}
The last term in the integrand of (\ref{bfec}) is related to
(\ref{P}-\ref{P0}). So, we can write
\begin{equation}
\label{fP2}
<K.P_{2}>:=\lim_{\varepsilon\to0}\int
dx^{4}K_{\mu}\Theta_{4}^{\mu\nu}(K_{\nu}-V_{\nu})\delta(\rho-\varepsilon),
\end{equation}
which is null because
$$\lim_{\rho\to0}\frac{R.\lbrace[R,V].\eta.[R,V]\rbrace.\eta.(R-V\rho)}{\rho^{6}}=0;$$
and
\begin{equation}
\label{fP1a}
<K.P_{1}>:=\lim_{\varepsilon\to0}\int
dx^{4}K_{\mu}{\Big\lbrace}\Theta_{3}^{\mu\nu}(K-V)_{\nu})+\Theta_{4}^{\mu\nu}K_{\nu}\rho\;{\hbox{\Large a}}_K{\Big\rbrace}\delta(\rho-\varepsilon),
\end{equation}
which is also null because
$$
\lim_{\rho\to0}\frac{1}{\rho^{5}}{R.\lbrace[R,V].\eta.[R,\rho\;\hbox{\Large
 a}+V\hbox{\Large a}.R].(R-\rho\;V)\rbrace}=
\lim_{\rho\to0}\frac{1}{\rho^{5}}{R.\eta.\lbrace{[R,V].\eta.[R,V]\rbrace}.R\hbox{\Large a}.R}=0,$$
as they can be easily verified. Finally,
\begin{equation}
\label{fP0a}
<K.P_{0}>:=\int
dx^{4}K_{\mu}{\Big\lbrace}\Theta_{2}^{\mu\nu}(K-V)_{\nu}+\Theta_{3}^{\mu\nu}K_{\nu}\rho\;{\hbox{\Large a}}_K{\Big\rbrace}\delta(\rho-\varepsilon)=0,
\end{equation}
as a consequence of (\ref{C}).

\newpage
\begin{center}
FIGURE 1 CAPTION
\end{center}
Fundamental process: the electron 4-velocity $V_{1}$ changes to  $V_{2}$ after
the emission/absorption of a classical photon of 4-velocity K.
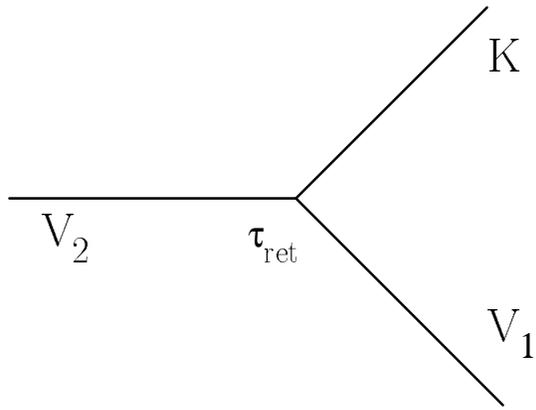
\end{document}